\documentclass{llncs}
\usepackage{amssymb}
\usepackage{amsmath}
\usepackage[english]{babel}
\usepackage{amsfonts, amsmath, xspace, tabularx, graphicx, xtab, ctable, verbatim, pifont, slashbox}
\usepackage[lofdepth,lotdepth]{subfig}

\newcolumntype{R}{>{\raggedleft\arraybackslash}X}%

    \usepackage{mathrsfs}

\newlength{\pageheight}
\newcommand{\ti}[1]{\textit{#1}\xspace}

\definecolor{brown}{RGB}{210, 105,30}

\newcommand{\ptep}{\ptep'\xspace}

\newcommand{\bcmp}{BCMP\xspace}

\begin{document}
    \mainmatter

    \title{Task Elimination may Actually Increase Throughput Time}
    \author{D.M.M. Schunselaar \and H.M.W. Verbeek}

    \institute{Eindhoven University of Technology,\\
    P.O. Box 513, 5600 MB, Eindhoven, The Netherlands\\
      \email{d.m.m.schunselaar@tue.nl, h.m.w.verbeek@tue.nl}\\
    }

    \maketitle

	\begin{abstract}
		The well-known Task Elimination redesign principle suggests to remove unnecessary tasks from a process to improve on time and cost. Although there seems to be a general consensus that removing work can only improve the throughput time of the process, this paper shows that this is not necessarily the case by providing an example that uses plain M/M/c activities. This paper also shows that the Task Automation and Parallelism redesign principles may also lead to longer throughput times. Finally, apart from these negative results, the paper also show under which assumption these redesign principles indeed can only improve the throughput time.
	\end{abstract}

	\section{Introduction}
	Within the Business Process Redesign community, a well-know redesign principle is the \ti{Task Elimination} redesign principle, e.g., see~\cite{Mansar:2005:BPB:1099116.1099120}. The idea behind this redesign principle is that, by removing work, the overall throughput time will decrease and that costs are reduced~\cite{Reijers2005283}. After all, cases which would have been processed by a particular task can now skip this task and immediately arrive at a subsequent task.
		
	Unfortunately, the redesign principle does not need to reduce the throughput time; it might actually \ti{increase}. Due to overtaking cases, some cases may become slower while the overtaking cases may not become faster. As a result, the average throughput time increases.
	
	 Within~\cite{goldratt2004the}, similar ideas are presented as the Task Elimination redesign principle, i.e., removal of bottlenecks to reduce the throughput time. Furthermore, within our earlier work~\cite{DBLP:conf/simpda/SchunselaarVRA14}, we have used, i.a., the Task Elimination redesign principle as a base line upon which we have created patterns to deduce if the throughput time of one process model is at-least-as-good as another process model. For all the aforementioned the same holds: an expected reduction of the throughput time might actually be an increase of throughput time. Within this paper, we sketch the boundaries as well as additional assumption under which the Task Elimination redesign principle will not increase the expected throughput time.
	
    The structure of this report is as follows: first, we touch upon some preliminaries. Second, we present an example process model on which Task Elimination does not reduce but \emph{increases} the throughput time. Third, we present sets of additional assumptions under which Task Elimination does reduce the throughput time. Fourth, we show in the discussion section that the reasoning for Task Elimination can also be applied to Task Automation and Parallelism. Finally, fifth, we present our conclusions. An extended version of this paper can be found in~\cite[Ch. 7]{Schunselaar2016}.
    
    \section{Preliminaries}
	Within this Section, we briefly introduce two modelling formalisms. The first, Petri nets~\cite{petriNets}, is used to show our motivating example why Task Elimination can result in an increased throughput time. The second, \bcmp networks, is used to guarantee under which assumptions Task Elimination cannot result in an increased throughput time.

    \subsubsection{Petri nets}    
\begin{figure}
    	\centering
    	\includegraphics[width=0.5\textwidth]{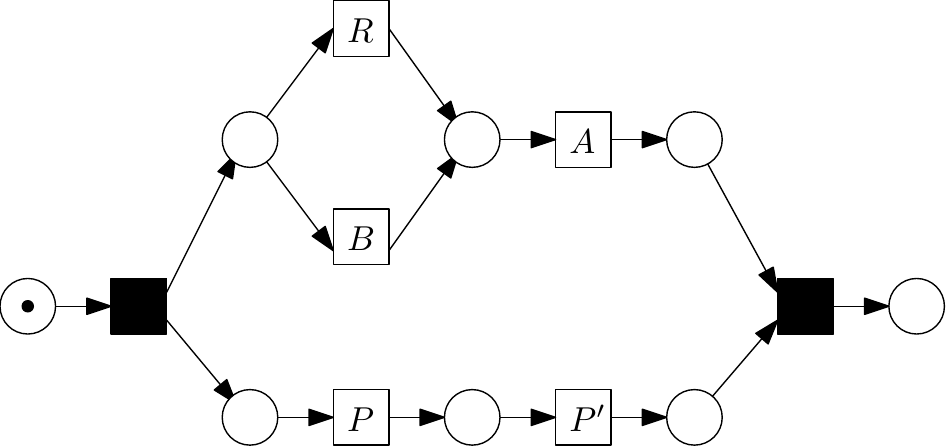}
    	\caption{Example Petri net.}
    	\label{fig:examplePetriNet}
    \end{figure}

    Petri nets were first introduced in~\cite{petriNets}. A Petri net consists of places (circles), transitions (rectangles), and edges between both (Fig.~\ref{fig:examplePetriNet}). Transitions denote the tasks of a Petri net and come in two flavours; labelled, and silent. Labelled transitions have a name, e.g., $A$. Silent transitions do not have a label and are usually indicated with a black rectangle.
    
    The behaviour of a Petri net is encoded using \ti{tokens}. Tokens are indicated by black dots. The state, or \ti{marking}, of a Petri net is the distribution of tokens over the places. A transition $t$ is allowed to \ti{fire} if all places in its \ti{preset} are marked, i.e., contain a token. The preset of a transition are those places that are the source of an edge to said transition. By firing a transition, a token from each of the preset places is \ti{consumed} and on each of the \index{Postset}\ti{postset} places a token is \ti{produced}. The postset of a transition are those places that are the target of an edge from said transition.
    
    We have two special states; the \ti{initial state} and the \ti{final state}. The initial state of a Petri net is the state from which we started firing transitions, e.g., in Fig.~\ref{fig:examplePetriNet}, the left-most place is an initial state. The final state is the state in which the Petri net is considered to be terminated, e.g., the right-most place in Fig.~\ref{fig:examplePetriNet}.
    
    \subsubsection{BCMP Networks}    
    \begin{figure}[tb]
    	\centering
    	\includegraphics[width=\textwidth]{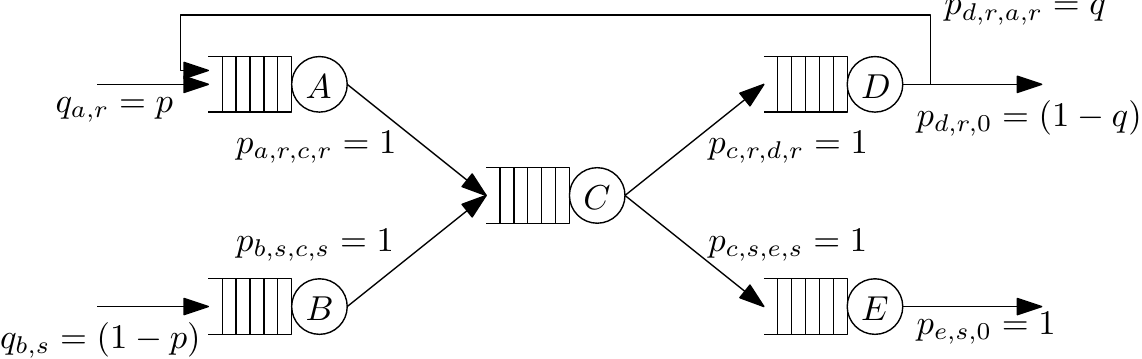}
    	\caption{Example queueing model.}
    	\label{fig:exampleQueueingModel}
    \end{figure}
    
    In order to explain \ti{queueing networks} and \index{BCMP}\ti{\bcmp} networks~\cite{DBLP:journals/jacm/BaskettCMP75} in particular, we start with the queueing model in Fig.~\ref{fig:exampleQueueingModel}. Where a Petri net has transitions, a queueing model has \ti{service centres}. In queueing networks, the term \ti{customer} is used to denote a token (although customers are not produced and consumed as in a Petri net). In its most simple form, customers arrive according to an exponential distribution with rate parameter $\lambda$. Customers receive service at a service centre by a \ti{server} with a service rate of $\mu$ (again exponentially distributed). The average time in between two arrivals is $\frac{1}{\lambda}$ and the average time needed to handle a customer is $\frac{1}{\mu}$. The resource utilisation is denoted by $\rho$ which in this case is equal to $\frac{\lambda}{\mu}$. If $\rho \geq 1$, then there is more work arriving per time unit than the service centre can handle. As such, in general, the requirement $\rho < 1$ is imposed.
    
    Within queueing networks, customers can be of different \ti{classes}. This would correspond to different types of tokens, e.g., a complaint by a gold customer or a complaint by a silver customer. Customers go from one service centre to the next service centre with a certain probability. These probabilities can depend on the customer class, e.g., in Fig.~\ref{fig:exampleQueueingModel}, after being processed at service centre $A$, a customer of class $r$ moves to the queue of service centre $C$ (again as a class $r$ customer). This probability is encoded as $p_{a, r, c, r}$. Note that customers can change class while moving between service centres. 
    
    Next to customer classes, queueing networks can contain \ti{feedback}, i.e., loops, and \ti{fork-join}, i.e., parallelism. Furthermore, the queueing network can be open (customers arrive and leave the network), closed (a fixed number of customers are in the network), or mixed (open for some classes of customer, and closed for other classes of customers). Within this paper, we only consider open queueing networks without fork-joins. To denote externally arriving customers and customers leaving the queueing network, the queueing model in Fig.~\ref{fig:exampleQueueingModel} has arrows without a source, or without a target. The incoming arrows, i.e., without a source, have a probability associated to them that a customer of a particular class will arrive at the service centre, e.g., the arrow with $q_{a, r} = p$ denotes that with probability $p$ a customer of class $r$ arrives at service centre $A$. Conversely, outgoing arrows have a probability that a customer of a particular class will leave the network after being processed at a particular service centre, e.g., $p_{d, r, 0} = (1 - q)$.
                  
    Within a \bcmp network, the service centres can be any of the following types~\cite{DBLP:journals/jacm/BaskettCMP75}:
    \begin{enumerate}
    	\item The service principle is First in First out. All customers receive the same service time distribution which is an exponential distribution.
    	\item There is a single server at a service centre. The service principle is server sharing. This means that every time unit every customer at this service centre receives a service rate proportional to the number of customers at this service centre. The service time distribution is arbitrary.
    	\item The number of servers in the service centre is greater than or equal to the maximum number of customers that can be queued.
    	\item There is a single server at a service centre, the queueing principle is preemptive-resume last-come-first-serve. Each class of customers may have a distinct distribution for the service times.
    \end{enumerate}
    Type $2, 3$ and $4$ service centres all assume service distributions with rational Laplace transforms. In~\cite{Cox1955}, for various distributions, it is shown they have rational Laplace transforms, e.g., exponential, hyperexponential, and hypoexponential. In the work of~\cite{DBLP:journals/acta/GelenbeM76}, it is stated that an arbitrary distribution may be as closely approximated as one may wish using a convex mixture of Erlang distributions. This mixture has a rational Laplace transform.
    
    One of the results from \bcmp networks is that, for \ti{open} networks where the \ti{Poisson} arrival of new customers is \ti{independent} of the state of the network, the number of customers in each service centre are \ti{independent} random variables~\cite{DBLP:journals/jacm/BaskettCMP75}. This means that the throughput time at a service centre is \ti{independent} of the other service centres.

    \section{Motivating example}
\begin{figure}
		\centering
		\includegraphics[width=\textwidth]{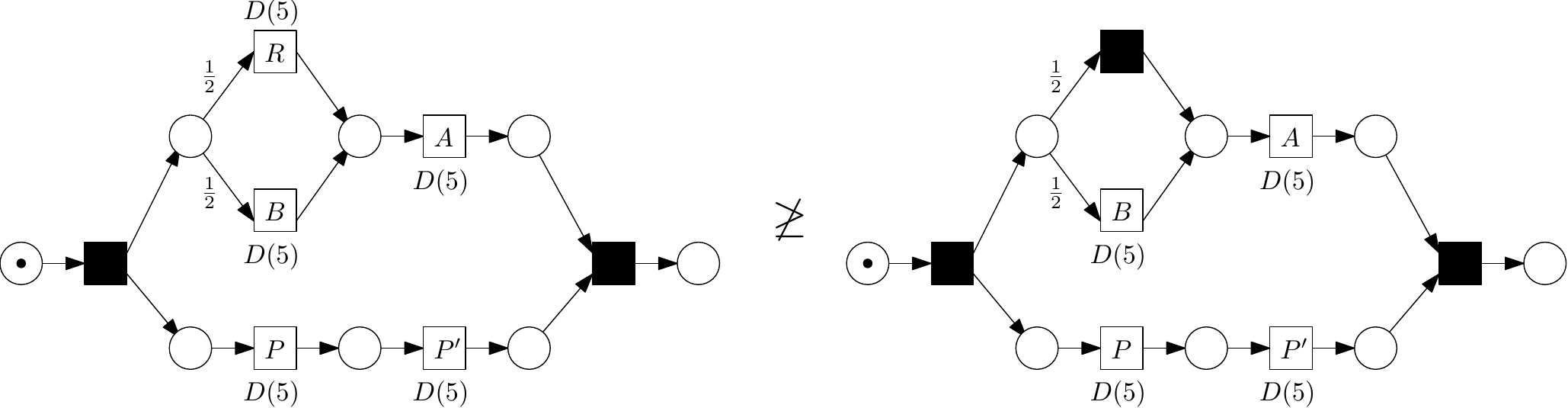}
		\caption{Example where \ti{Task Elimination} increases the throughput time.}
		\label{fig:CounterExampleParallelism}
	\end{figure}

    Figure~\ref{fig:CounterExampleParallelism}~\cite{Schunselaar2016} shows our example process model at the left-hand side, and the process model after applying the Task Elimination design principle on the right-hand side. As one can see, in the right-hand model, task $R$ has been eliminated. As a result, one would expect that the right-hand model would have a throughput time which is at-most that of the left-hand model. However, as we will show, this is not the case.

In the example process models, we have two types of cases: \ti{red} cases (executing task $R$, and skipping task $B$) and \ti{blue} cases (executing task $B$). Task $A$ is executed for \ti{all} cases. Finally, tasks $P$ and $P'$ are \ti{parallel} to the aforementioned tasks. For now, assume the processing time distribution of each task is \ti{deterministic} with the value as indicated in Fig.~\ref{fig:CounterExampleParallelism}, e.g., task $A$ takes 5 time units. Furthermore, assume the cases arrive in a \ti{burst}, i.e., within a very short period of time a large amount of cases arrive. Thereafter, for a long period of time, no cases arrive. Let the time between the bursts be such that the model is empty when a new burst arrives. During a burst, a total of $N$ cases arrive. Assume that $\frac{1}{2}N$ red cases and $\frac{1}{2}N$ blue cases arrive. Furthermore, assume that all tasks process the cases in a First in First out (FiFo) order.

	In the original process model, in the parallel branch, tasks $P$ and $P'$ process the cases in the order they arrive at the process model. At the tasks $R$ and $B$, the cases are also processed in the order they arrive. At task $A$, the ordering can be slightly different, i.e., two consecutively arriving cases can both be blue cases and then a red cases can be in between at $A$. In general, the order of the cases at $A$ is similar to the arrival of cases to the process model. Which in turn is similar to the order of the cases at $P$ and $P'$. As a result, the time cases spend on synchronising is relatively small at the transition joining the parallel branches (in the remainder, this transition is called: AND-join).

	In the redesigned process model, we have eliminated task $R$. This means that just after a burst has occurred, the queue at $A$ is filled with $\frac{1}{2}N$ red cases. Behind these red cases, the blue cases are enqueued as these first have to be processed by task $B$. In the parallel branch nothing has changed, i.e., tasks $P$ and $P'$ still process the cases in the order they arrive at the process model. This means that at the AND-join, the top branch has been resorted on red first and blue last, while in the bottom branch, no resorting was done and the cases arrive in the same order as they arrived at task $P$. Only after the $\frac{1}{2}N$ red cases have been processed by task $A$, blue cases can be synchronised.

	By eliminating task $R$, the sorting of cases before the AND-join has changed dramatically in the upper branch, while it did not change in the lower branch. As a result, task $A$ started processing cases which had no priority, as in the bottom branch, no work has been done on them. As such, there was no need to work on these cases by task $A$. This while cases which were being processed at tasks $P$ and $P'$ (blue cases) were at the end of the queue of task $A$. The \ti{overtaking} of cases makes that at the synchronisation the \ti{wrong} cases are waiting to be synchronised, i.e., cases which cannot be synchronised with any of the cases at the other branch.

	In our example, we have used burst arrivals and deterministic processing times. If we look into a \ti{Poisson} arrival stream and \ti{exponential distributions} on the tasks, i.e., not deterministic, then one can still observe the same phenomenon. Within a \ti{Poisson} arrival stream, it is still possible that multiple cases arrive within a short period of time. The earlier sketched overtaking of cases might be less prominent (there are fewer cases at the same time within the process model) but the end effect will be the same, i.e., Task Elimination can have a \ti{negative} effect on the throughput time.

	\begin{figure}[h!]
	\centering
	\includegraphics[width=\textwidth]{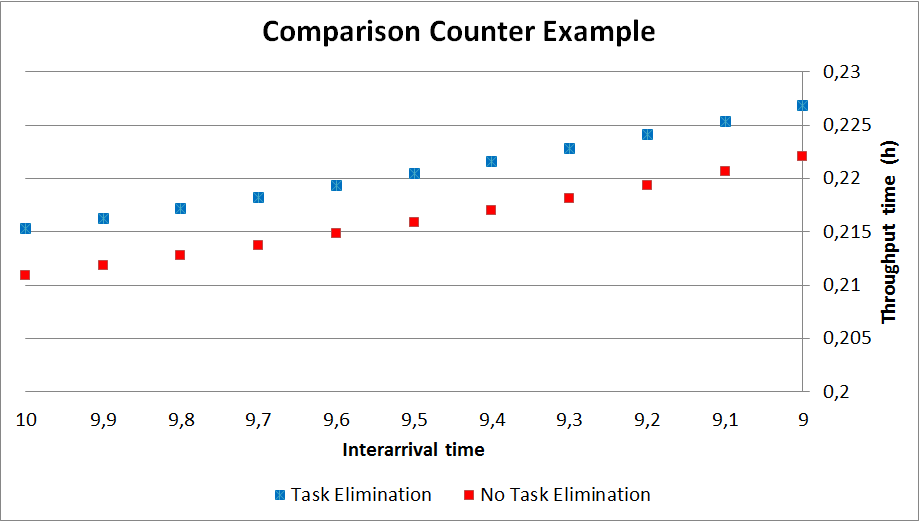}
	\caption{Quantitative results for our example. The arrival process is Poisson and the processing times are deterministic 5 minutes with FiFo queues at every task. The blue line is with Task Elimination and the red line is no Task Elimination.}
	\label{fig:QuantitativeDataCounterExample}
	\end{figure}

	We have simulated both models from Fig.~\ref{fig:CounterExampleParallelism} with different Poisson arrival distributions (interarrival times as shown in Fig.~\ref{fig:QuantitativeDataCounterExample}) with 30 replications and a replication length of 1000 hours using L-SIM~\cite{LSIM}. All tasks have a deterministic processing times of 5 minutes. Furthermore, every task has a FiFo queue. Finally, every task has exactly 1 dedicated resource allocated. The results are depicted in Fig.~\ref{fig:QuantitativeDataCounterExample} (blue is Task Elimination and red is no Task Elimination). One can indeed observe that \ti{Task Elimination} has a negative effect on the throughput time.
	
	\section{Additional assumptions}
	The above example shows the need for additional assumptions under which Task Elimination gives an intuitive result. Within this Section, we assume there are two models $M_1$ and $M_2$ where $M_2$ is obtained from $M_1$ by applying the Task Elimination design principle on some tasks (possibly none), i.e., these tasks have become silent transitions.
	
	
	
	\subsection{Scheduling}
\begin{figure}[h!]
		\centering
		\includegraphics[width=\textwidth]{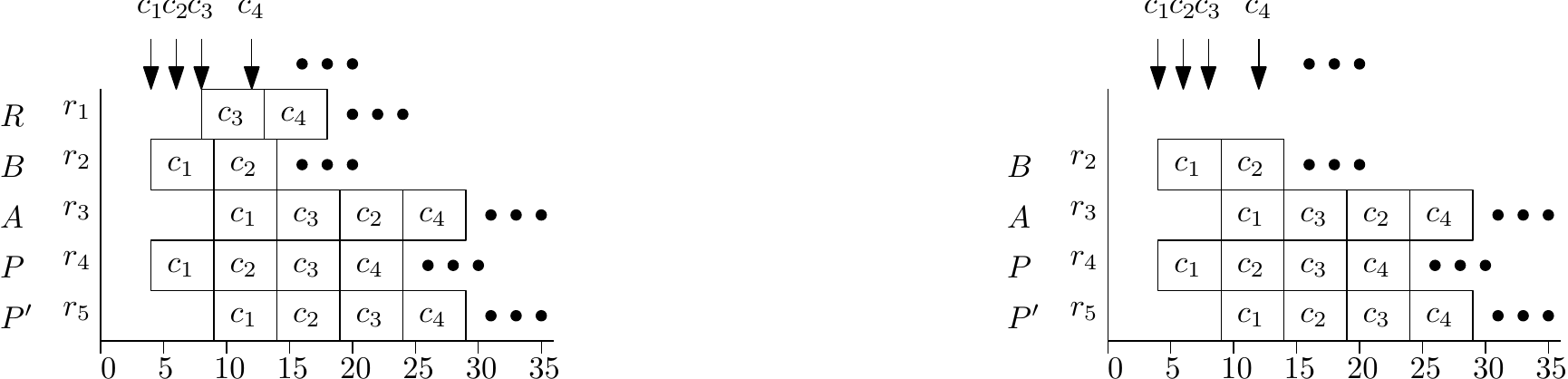}
		\caption{Example execution schedule for the Petri nets in Fig.~\ref{fig:CounterExampleParallelism}. On the y-axis, we have the tasks with the resources ($r_i$). On the y-axis, we have the time. The $c_i$s are the cases and the small downward arrows are the arrival of cases.}
		\label{fig:orderingTheSpaceOfModelsExampleScheduleMMp}
	\end{figure}

	In this set of assumptions~\cite{Schunselaar2016}, we assume that both process models follow the same schedule in executing the cases (see Fig.~\ref{fig:orderingTheSpaceOfModelsExampleScheduleMMp} for an illustration of an example schedule). This means that, even though a case $c'$ arrives earlier at a particular task than case $c$, the execution of $c'$ is possibly delayed to work on $c$ first (even though $c$ might arrive later at the respective task than $c'$). This setting is particularly applicable in situations where a fixed set of cases needs to be processed during a certain time period, e.g., order picking, and if a-priori a schedule is made to optimally execute the cases. It is easy to show that the redesign principles and conversely the patterns in~\cite{DBLP:conf/simpda/SchunselaarVRA14} can safely be applied. This follows from the fact that by eliminating tasks in the execution of cases, it is always possible to project the schedule for executing the cases of a model with more tasks to a model with fewer tasks. 

	\subsection{BCMP networks}
	For \bcmp networks~\cite{DBLP:journals/jacm/BaskettCMP75}, exact solutions exist to compute the mean throughput time. If a Petri net can be transformed to a \bcmp network, then we can show that \ti{Task Elimination} indeed returns the correct result. This follows from the fact that the throughput time at a task is \ti{independent} of the other tasks. By having fewer tasks, the throughput time of the cases does not increase. As a result, the throughput time of the process model will not increase and probably will decrease. A Petri net can be transformed to a \bcmp network if it is a state machine, i.e., every transition has exactly one place in its preset and exactly one place in its postset. Furthermore, the tasks within a Petri net need to adhere to the types supported by the \bcmp network~\cite{Schunselaar2016}. Finally, the arrival process of new cases needs to be \ti{Poisson}.
	
\begin{figure}[h!]
		\centering
		\includegraphics[width=\textwidth]{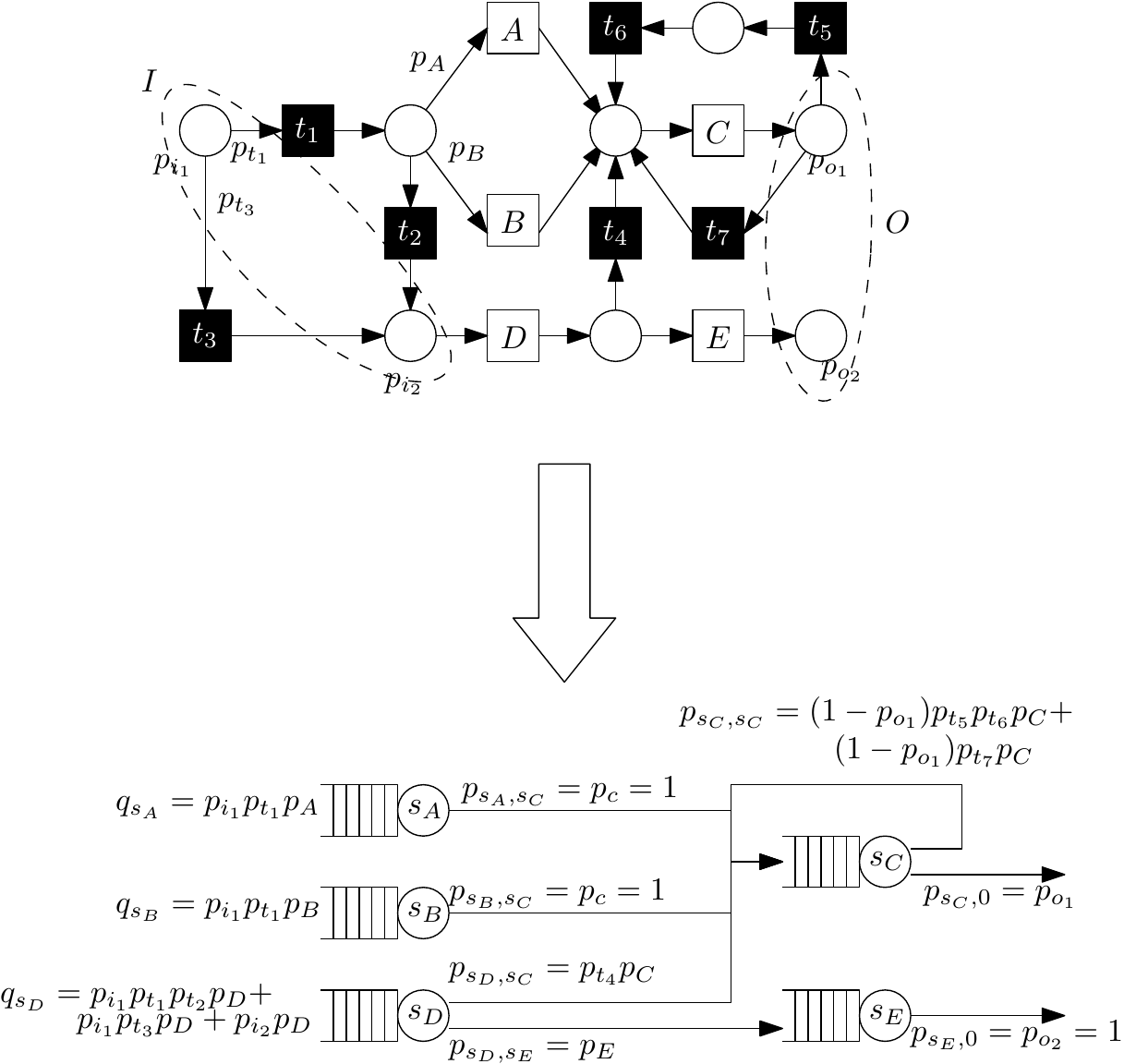}
		\caption{Example transformation of a state machine (with some probabilities made explicit) to a \bcmp network.}
		\label{fig:SMToBCMP}
	\end{figure}

	The transformation of a state machine to a \bcmp network is relatively straightforward (see Fig.~\ref{fig:SMToBCMP} for an example). As stated, a \bcmp network consists of service centres, an arrival rate, and probabilities for customers moving through the \bcmp network, i.e., arriving customers, customer moving between service centres, and leaving customers. Within Fig.~\ref{fig:exampleQueueingModel}, we denote with $p_{a, s, b, s}$ the probability of a customer moving from service centre $a$ to service centre $b$ with customer class $s$. For the transformation, we do not need to use the customer class for routing customers, hence we omit the customer class in our probabilities, i.e., we write $p_{a, b}$ to denote the probability of a customer moving from service centre $a$ to service centre $b$. 
	
	Within our transformation, since we only consider open queueing networks, we assume there is a set of places $I$ where customers/tokens are entering the Petri net, and a set of places $O$ where customers/tokens are leaving the Petri net. Furthermore, since the queueing network is open, every customer/token produced in a place in $I$ can always reach a place in $O$. Furthermore, we assume that every edge in the Petri net has a probability associated to it (in Fig.~\ref{fig:SMToBCMP}, we have illustrated some of the probabilities). The probability on the outgoing edge of a transition is by definition 1, the probability of the incoming edge of a transition $t$ is the probability that the transition will fire if there is a token in its preset (this probability is indicated by $p_t$). Next to this, a place $o \in O$ has a probability that a token will leave the Petri net from $o$ (indicated by $p_o$), and a place $i \in I$ have a probability that a token is produced in $i$ (indicated by $p_i$). We assume that the sum of all $p_i$ is equal to 1, e.g., $p_{i_1} + p_{i_2} = 1$ in Fig.~\ref{fig:SMToBCMP}. For every set $T$ of transitions that have the same preset, we have $\sum_{t \in T} p_t = 1$, e.g., $p_{t_1} + p_{t_3} = 1$ in Fig.~\ref{fig:SMToBCMP}.
	
	Every labelled transition in the Petri net is transformed into a service centre with its respective distribution, we indicate with $s_t$ the service centre of transition $t$. The arrival probability of a new customer at service station $s_t$ (indicated by $q_{s_t}$ in Fig.~\ref{fig:SMToBCMP}) is 0 if the preset of $t$ is not in $I$ and $p_i p_t$ if it is. Mutatis mutandis, we can deduce the probability that a customer leaves the network after having received service at a service station. The probability of a customer arriving at service station $s_{t'}$ after having received service at service centre $s_t$ (indicated by $p_{s_t, {s_t'}}$ in Fig.~\ref{fig:SMToBCMP}) is $p_{t'}$ if the preset of $t'$ equals the postset of $t$ (note that we have a state machine so every transition only has a single place in its preset/postset). In case of silent transitions, there might be a path of silent transitions between the postset of $t$ and the preset of $t'$. In this case, the probability $p_{s_t, {s_t'}}$ is equal to the sum of the probabilities of all possible paths between $t$ and $t'$, e.g., from transition $C$ to transition $C$ in Fig.~\ref{fig:SMToBCMP}. The probability of a path is equal to the product of the probabilities of the edges in a path. If no such path exists, then $p_{s_t, {s_t'}} = 0$ (implicit in Fig.~\ref{fig:SMToBCMP}).
	
	Using the above sketched transformation, we can transform a state machine into a \bcmp network. Hence, Task Elimination will not result in an increase of throughput time within state machines that can be transformed.
	
\section{Discussion}
We have focussed primarily on Task Elimination. These results are also applicable to the \ti{Task Automation}~\cite{Mansar:2005:BPB:1099116.1099120} and \ti{Parallelism} redesign principles~\cite{Reijers2005283}. The Task Automation redesign principle essentially reduces the the time spent in a task to almost 0. As a result, the same reasoning as for Task Elimination can be followed. Within the Parallelism redesign principle, multiple tasks are put in parallel to decrease the throughput time. If we take our example process model in Fig.~\ref{fig:CounterExampleParallelism} and we would replace $R$ by the sequential execution of $R_1, \ldots, R_n$ such that duration of $R$ is the same as the sum of the durations of $R_1, \ldots, R_n$ and the duration of $R_1, \ldots, R_n$ is close to 0, then by placing $R_1, \ldots, R_n$ in parallel, we possibly would make that branch significantly faster. As a result, in front of $A$, we would have only red cases enqueued.

\section{Conclusion}
Within this paper, we have shown that various redesign principles and in particular \ti{Task Elimination} does not need to reduce the throughput time under all circumstances; under some circumstances, the throughput time actually \ti{increases}. As a result, our patterns presented in~\cite{DBLP:conf/simpda/SchunselaarVRA14} do not always yield the correct result. At the same time, we have presented additional assumptions/settings under which Task Elimination always does give the correct result.

    \bibliographystyle{splncs}
    \bibliography{main}

\end{document}